\documentclass[showpacs,showkeys,prb,preprint,aps,superscriptaddress,preprintnumbers,amsmath,amssymb]{revtex4}
\usepackage{graphicx}
\usepackage{dcolumn}
\usepackage{bm}
\usepackage[dvips]{color}
\begin{document}
%
%
\title{EuNiGe$_3$, an anisotropic antiferromagnet}
\author{Arvind Maurya}
\affiliation{Department of Condensed Matter Physics and Materials
Science, Tata Institute of Fundamental Research, Homi Bhabha Road,
Colaba, Mumbai 400 005, India}
\author{P. Bonville}
\affiliation{CEA, Centre d'Etudes de Saclay, DSM/IRAMIS/Service de Physique de I'Etat Condens\'e, 91191 Gif-sur-Yvette, France}
\author{A. Thamizhavel}
\affiliation{Department of Condensed Matter Physics and Materials
Science, Tata Institute of Fundamental Research, Homi Bhabha Road,
Colaba, Mumbai 400 005, India}
\author{S. K. Dhar}
\email{sudesh@tifr.res.in}
\affiliation{Department of Condensed Matter Physics and Materials
Science, Tata Institute of Fundamental Research, Homi Bhabha Road,
Colaba, Mumbai 400 005, India}
\date{\today}
%
\begin{abstract}
Single crystals of EuNiGe$_3$ crystallizing in the non-centrosymmetric BaNiSn$_3$-type structure have been grown using In flux, enabling us to explore the anisotropic magnetic properties which was not possible with previously reported polycrystalline samples. The EuNiGe$_3$ single crystalline sample is found to order antiferromagnetically at 13.2\,K as revealed from the magnetic susceptibility, heat capacity and electrical resistivity data. The low temperature magnetization M(H) is distinctly different for field parallel to \textit{ab}-plane and \textit c-axis; the \textit{ab}-plane magnetization varies nearly linearly with field before the occurrence of an induced ferromagnetic phase (spin-flip) at 6.2\,Tesla; on the other hand M(H) along the \textit c-axis is accompanied by two metamagnetic transitions followed by a spin-flip at 4.1\,T. A model including anisotropic exchange and dipole-dipole interactions reproduces the main features of magnetization plots but falls short of full representation. (H,T) phase diagrams have been constructed for the field applied along the principal directions. From the $^{151}$Eu M\"{o}ssbauer spectra, we determine that the 13.2\,K transition leads to an incommensurate antiferromagnetic intermediate phase followed by a transition near 10.5\,K to a commensurate antiferromagnetic configuration.
\end{abstract}

%
\maketitle
\section{Introduction}
Determining the magnetic structure of divalent Eu compounds remains a challenge since neutron diffraction is not easy to perform in such materials due to the large absorption of neutrons by Eu isotopes. One way of obtaining information about the spin arrangement is to perform single crystal magnetization measurements. M\"{o}ssbauer spectroscopy on the isotope $^{151}$Eu in polycrystalline samples can also yield information about the commensurability of the magnetic structure with the lattice. Recently, we studied single crystals of two members of the EuMX$_3$ family, where M~=~Pt and X is Ge or Si,~\cite{Neeraj EuPtSi3, Neeraj EuPtGe3} which crystallize in the body-centered tetragonal BaNiSn$_3$-type structure which is noncentrosymmetric (space group \textit{I4mm}). The properties of a polycrystalline sample of a third compound of the family, EuNiGe$_3$, have recently been reported in the literature.~\cite{Goetsch} EuNiGe$_3$ orders antiferromagnetically at T$_N$=13.6\,K. Assuming a collinear A-type antiferromagnetic (AFM) structure in which ferromagnetic layers of Eu magnetic moments in the \textit {ab}-plane alternate along the c-axis with AFM coupling, the in-plane J$_1$ and interplane J$_c$ nearest neighbor exchange integrals were derived in Ref.~\onlinecite{Goetsch}.

In the present work, we report on the synthesis of a single crystal of EuNiGe$_3$. We have measured the magnetization along the principal directions of the crystal and applied $^{151}$Eu M\"{o}ssbauer spectroscopy to study the hyperfine interaction. The heat capacities in zero and applied magnetic field, electrical resistivity and magnetoresistivity of the single crystal have also been measured. The prominent feature of our data is the magnetization curves at 1.8\,K, which are strongly anisotropic, the curve along the [001] crystal axis showing a metamagnetic staircase-like behavior. We present a model, including the dipole-dipole interaction and exchange anisotropy, to interpret these findings.
\section{Experimental Details}
The single crystal of EuNiGe$_3$ was grown by using a high temperature solution growth method with In as a solvent. Initially we used Sn flux as it had resulted earlier in the successful growth of the single crystals of EuPtX$_3$ (X~=~Si or Ge),~\cite{Neeraj EuPtSi3, Neeraj EuPtGe3} but we failed to obtain the single crystals of the Ni compound. However, we were successful in growing the single crystals of EuNiGe$_3$ from In solvent. Arc melted polycrystalline EuNiGe$_3$ and excess indium kept inside an alumina crucible were sealed in vacuum inside a quartz ampoule. The temperature of the ampoule was uniformly raised to 1100$^{\circ}$C/h in 24~hours and held at that value for another 24~hours to ensure homogeneity of the solution. After this a cooling rate of 2$^{\circ}$C/h was employed upto 600$^{\circ}$C followed by relatively faster cooling (60~$^{\circ}$C/h) to room temperature. The excess indium flux was removed by centrifugation process. The quality of the single crystals was confirmed by recording the Laue patterns which showed sharp spots conforming to the tetragonal symmetry. A portion of the crystal was powdered for x-ray diffraction; all the peaks of the spectrum could be indexed to the tetragonal structure with space group \textit{I4mm} (\#107). The absence of any extra peaks confirms the phase purity of the grown crystal. The single crystal was cut along the desired directions by electric spark discharge under a suitable dielectric and the orientation of the desired plane(s) confirmed by Laue patterns. Magnetization was measured in a Quantum Design VSM magnetometer, and the heat capacity and electrical resistivity were measured in a Quantum Design PPMS. $^{151}$Eu M\"{o}ssbauer spectra were recorded at a few selected temperatures using a constant acceleration spectrometer with a $^{151}$Sm$^*$F$_3$ source.
\section{Results}
\subsection{Magnetic properties}
The Rietveld analysis of the powder x-ray diffraction spectrum yielded the lattice parameters \textit{a}~=~0.4338(8)\,nm and \textit{c}~=~0.9895(9)\,nm, in good agreement with the previously reported values.

The inverse susceptibility, $\chi^{\rm {-1}}$, of EuNiGe$_3$ is practically isotropic above 20\,K for the three directions [100], [110] and [001] of the magnetic field (0.1\,T). The high temperature susceptibility and the fit to the Curie-Weiss expression $\chi$~=~C/(T $- \theta_{\rm p}$) in the range 50~-~300\,K yields the following results: $\mu_{\rm eff}$~=~7.89, 7.87 and 7.90\,$\mu_{\rm B}$/Eu and $\theta_{\rm p}$~=~3.4, 3.8 and 5.1\,K along [100], [110] and [001], respectively. The magnitude of $\mu_{\rm eff}$ is consistent with the divalent state of the Eu ions ($\mu_{\rm eff}$~=~7.94\,$\mu_{\rm B}$/Eu$^{2+}$) and the polycrystalline average 4.1\,K of $\theta_{\rm p}$ is comparable to the value (5\,K) reported previously in Ref.~\onlinecite{Goetsch}. 
%
\begin{figure}[h]
\includegraphics[width=0.45\textwidth]{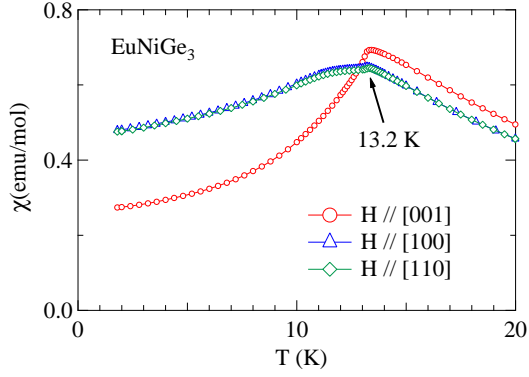}
\caption{\label{Chi_vs_T_low_temp_all_directions}(Color online) Low temperature part of the magnetic susceptibility $\chi$=M/H in EuNiGe$_3$ for a field of 0.1\,T applied along [001], [100] and [110].}
\end{figure}
%
A similar situation of positive $\theta_{\rm p}$ along [100] and [001] was also encountered in EuPtSi$_3$~\cite{Neeraj EuPtSi3} which undergoes two AFM transitions at 17 and 16\,K, respectively. It was tentatively attributed to an AFM nearest neighbor exchange, J$_1$, and a ferromagnetic (F) next neighbor exchange J$_2$ with $J_2>\vert J_1\vert$. The susceptibility M/H below 20\,K along the three directions is shown in Fig.~\ref{Chi_vs_T_low_temp_all_directions}. The antiferromagnetic transition is marked by a peak at $T_{\rm N}$~=~13.2\,K. The transition temperature is slightly lower than that reported on the polycrystalline sample. The susceptibility is isotropic in the \textit{ab}-plane but its magnitude at $T_{\rm N}$ is distinctly higher for $H~\parallel$~[001]. This difference, reflected in the higher value of $\theta_p$ along [001], persists for T~$>T_{\rm N}$ but decreases with temperature and eventually disappears above 45\,K. It is due to the crystal field and exchange anisotropy to be introduced below in section \ref{model}.
The susceptibility along the \textit{c}-axis [001] at 1.8\,K is 0.27\,emu/mol and is field-independent for applied fields of $H$~=~0.005, 0.02 and 0.1\,T, in the temperature range below 20\,K. The \textit{ab}-plane susceptibility at 1.8\,K in a field of 0.1 T is about twice larger with a value of 0.47\,emu/mol. 
\begin{figure}[h]
\includegraphics[width=0.45\textwidth]{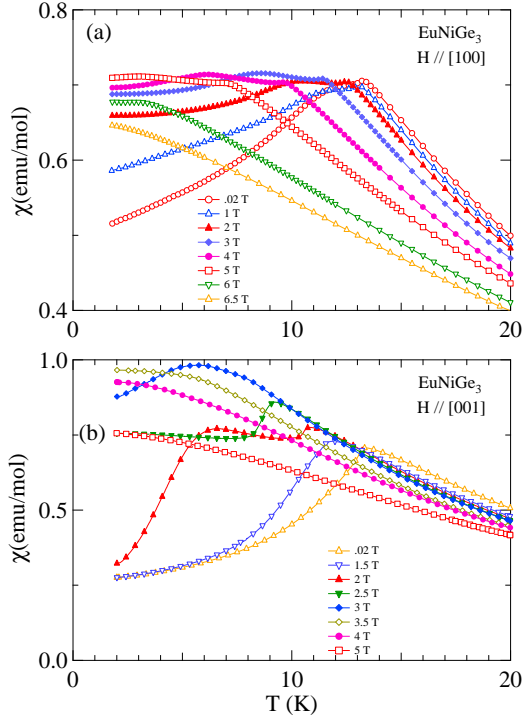}
\caption{\label{fig_Chi_vs_T_100_diff_fields}(Color online) Low temperature susceptibility for (a) $H~\parallel$~[100] and (b) $H~\parallel$~[001] in EuNiGe$_3$ for different values of the applied field.}
\end{figure}
The evolution of magnetic susceptibility of EuNiGe$_3$ with magnetic field applied along [001] and [100] at low temperatures is shown in Fig.~\ref{fig_Chi_vs_T_100_diff_fields}. There is a broad hump in the \textit{ab}-plane susceptibility centered near $T_{\rm m}$=10.5\,K and the temperature dependent magnetization shows a field dependence even at low fields (see Fig.~\ref{fig_Chi_vs_T_100_diff_fields}), unlike the data for $H~\parallel$~[001]. However, the field dependence of the susceptibility is seen even for $H~\parallel$~[001] at higher fields as inferred from Fig.~\ref{fig_Chi_vs_T_100_diff_fields})b. The broad hump is presumably the phase boundary between commensurate and incommensurate modulated antiferromagnetic structures, which is elaborated in the M\"{o}ssbauer spectroscopy section.The $T_{\rm N}$ decreases with increase in the strength of the magnetic field as expected for an antiferromagnetic compound.  Ref.~\onlinecite{Goetsch} reports a cusp in the susceptibility at 5\,K in low fields ($<$~500\,Oe), but such a feature is absent from our data. \\

\begin{figure}[h]
\includegraphics[width=0.45\textwidth]{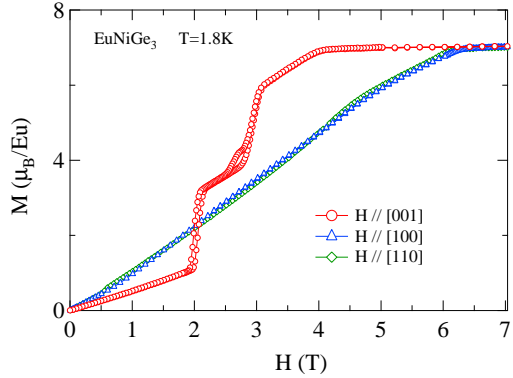}
\caption{\label{fig_M_vs_H_1p8K_all_directions}(Color online) Isothermal magnetization curves at 1.8\,K in EuNiGe$_3$ for  field along [100], [110] and [001].}
\end{figure}

\begin{figure}[h]
\includegraphics[width=0.45\textwidth]{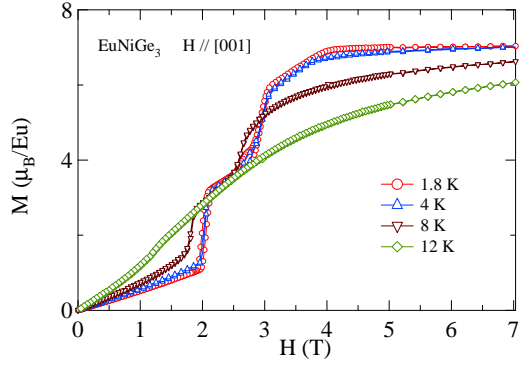}
\caption{\label{fig_M_vs_H_001_diff_temperatures}(Color online) Magnetization curves along [001] in EuNiGe$_3$ at different temperatures.}
\end{figure}

The magnetization M(H) at 1.8\,K along the three directions [100], [110] and [001] is shown in Fig.~\ref{fig_M_vs_H_1p8K_all_directions}.
For {\bf H} in the \textit{ab}-plane, M varies almost linearly with the field up to a spin-flip (or saturation) field of 6.2\,T, with a slight slope change at 4.6\,T. In antiferromagnets, such a response is typically seen when the field lies along a hard axis or inside a hard plane. So the Eu moments of the zero field magnetic structure do not lie in the (\textit{ab}) plane in EuNiGe$_3$; they could be perpendicular to this plane.
The behavior of the magnetization is much more remarkable when the field is applied along [001].  It first increases linearly with the field, then undergoes a spin-flop like jump at 2\,T followed by another spin-flop like feature near 3\,T, and by a spin-flip transition at 4.1\,T. This suggests that [001] is the easy axis of magnetization. However, the [001] magnetization is not zero below the first spin-flop field of 2\,T, indicating that the magnetic structure is more complicated than a simple collinear bipartite AF structure. The saturation magnetization along all the three directions reaches 7\,$\mu_{\rm B}$/f.u., matching with the theoretical T~=~0 spin only moment of Eu$^{2+}$ (S~=~7/2 and L~=~0). The observed difference in spin-flip field values is rather unexpected in a material where crystalline anisotropy is weak, and this behavior could be due to exchange anisotropy or arise from dipole-dipole interactions. In section \ref{model}, we present a model which tries to account for this interesting behavior, without full success however. The M(H) curves for $H~\parallel$~[001] at selected temperatures are shown in Fig.~\ref{fig_M_vs_H_001_diff_temperatures}. The staircase like behavior is getting blurred as temperature increases, with decreasing spin-flop fields.
%
%
\subsection {Heat Capacity}
\begin{figure}[h]
\includegraphics[width=0.45\textwidth]{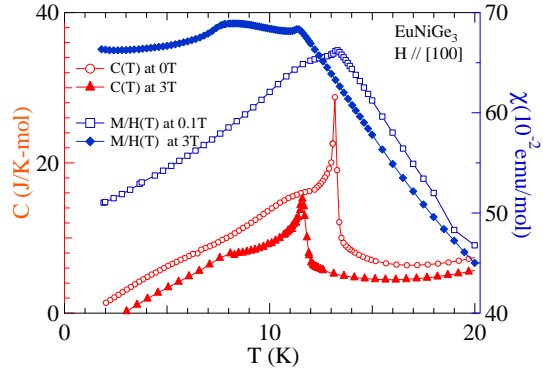}
\caption{\label{fig_Chi_and_HC_comparison}(Color online) Heat capacity (left scale) and magnetic susceptibility (right scale) in EuNiGe$_3$ for fields of 0 and 3\,T. The 3\,T heat capacity curve is shifted down by 3\,J/mol.K for clarity.}
\end{figure}

The heat capacity between 1.8 and 30\,K in zero magnetic field and with a field of 3\,T is plotted in Fig.~\ref{fig_Chi_and_HC_comparison}. A single sharp peak  at 13.2\,K in the zero-field data is representative of the AFM transition; a broad anomaly centered around 10.5\,K correlates well with the anomaly in the susceptibility for $H~\parallel$~[100] observed near the same temperature. The correlation holds in an applied magnetic field as well, at least upto 3\,T. For this field value, the narrow peak has shifted down to 11.6\,K and the broad anomaly occurs near 8\,K.
For the sake of comparison, the M/H data measured at the same value of the field for $H~\parallel$~[100] are also plotted in Fig.~\ref{fig_Chi_and_HC_comparison}. There is a very good correspondence between the two sets of the data. The zero-field $^{151}$Eu M\"ossbauer spectra, to be described in section \ref{mossb}, reveal that the  transition  at T$_{\rm N}$=13.2\,K leads to an incommensurate (ICM) AFM phase, and that the anomaly near 10.5\,K marks a transition to a commensurate AFM phase. This latter transition temperature will be referred to in the following as T$_{\rm IC}$.
 \subsection{Resistivity and magnetoresistivity}
\begin{figure}[t]
\includegraphics[width=0.45\textwidth]{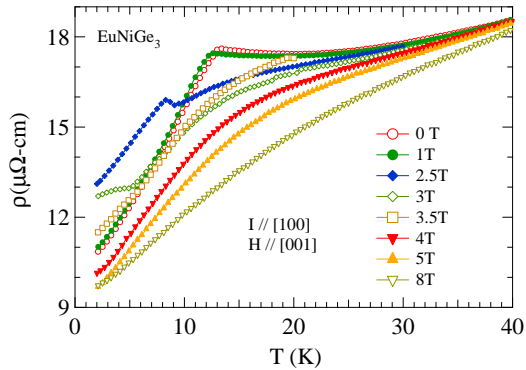}
\caption{\label{fig_RT_I_100_H_001}(Color online) Thermal variation of the resistivity in EuNiGe$_3$ for a current along [100] and various values of the field applied along [001].}
\end{figure}

The electrical resistivity $\rho$(T) of EuNiGe$_3$ with the current density $J~\parallel$~[100] in zero field and for different values of the field applied along [001] is shown in Fig.~\ref{fig_RT_I_100_H_001}. Qualitatively, its thermal variation is similar to that reported for the polycrystalline sample in Ref.~\onlinecite{Goetsch}, with an anomaly at the AFM transition temperature.
The residual resistivity ratio of our single crystal sample, defined as R(300\,K)/R(1.8\,K), is worth 5.5 and is thus almost an order of magnitude smaller than that of the polycrystal in Ref.~\onlinecite{Goetsch}. The values of R(300\,K)/R($\simeq T_N$) are however comparable: 3.5 for the single crystal compared to 7 for the polycrystalline sample. It may be recalled that there is an anomaly near 5\,K in the susceptibility of the polycrystalline sample, which is absent in the single crystal. Though the resistivity value at 300\,K in the two samples is nearly the same, the polycrystalline sample has a resistivity of 1.5\,$\mu\Omega$cm at 1.8\,K, compared to 11\,$\mu\Omega$cm for the single crystal.
\begin{figure}[h]
\includegraphics[width=0.45\textwidth]{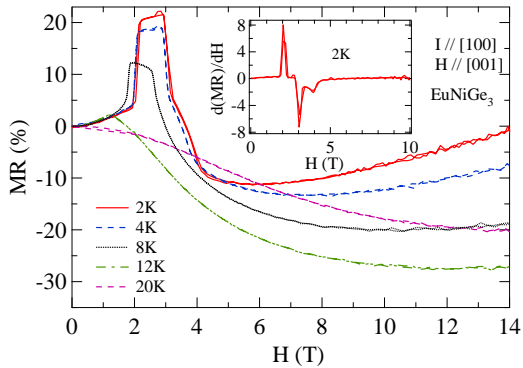}
\caption{\label{fig_MR_001_diff_temperatures}(Color online) Magnetoresistance with $H~\parallel$~[001] in EuNiGe$_3$ at various temperatures. The inset shows the field derivative of the MR ratio at 2\,K.}
\end{figure}
In the presence of the magnetic field, the magnetic transition temperature can be easily identified as the resistivity decreases rapidly below $T_{\rm N}$. As seen in Fig.~\ref{fig_RT_I_100_H_001}, application of a field decreases $T_{\rm N}$, which is consistent with the magnetization and the heat capacity data discussed above. At H~$\geq$~3.5\,T the anomaly in the resistivity due to the magnetic transition is no longer discernible. For H~$\geq$~4\,T the resistivity is less than the corresponding zero-field values in the range 1.8-40\,K shown in the figure. Typically the magneto-resistance ratio MR, defined as [R(H)-R(0)]/R(0), of an ordinary polycrystalline metal is positive and of the order of a few \%. MR can be appreciable for metals and single crystals with large residual resistivity ratio which is not the case with our sample.
\begin{figure}[h]
\includegraphics[width=0.45\textwidth]{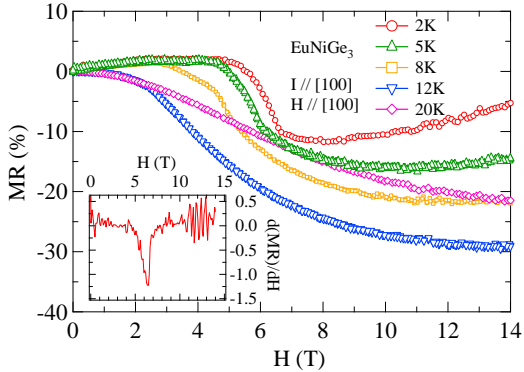}
\caption{\label{fig_MR_100_diff_temperatures}(Color online) Magnetoresistance with $H~\parallel$~[100] in EuNiGe$_3$ at various temperatures. The inset shows the field derivative of the MR ratio at 2\,K.}
\end{figure}
A negative MR suggests a magnetic origin and indicates that short range interactions, suppressed by the field, are present above $T_{\rm N}$. At lower fields regions of both positive and negative MR are observed and a monotonically varying behavior is not present. The MR at a few selected temperatures measured up to 14\,T is shown in Fig.~\ref{fig_MR_001_diff_temperatures}. The field variation of MR is tightly correlated with that of the magnetization for $H~\parallel$~[001] at various temperatures (Fig.~\ref{fig_M_vs_H_001_diff_temperatures}). At 2, 4 and 8\,K the sharp, almost vertical upturn in MR occurs at the value of the field where the first metamagnetic transition takes place, 2\,T at 1.8\,K.
%
\begin{figure}[h]
\includegraphics[width=0.45\textwidth]{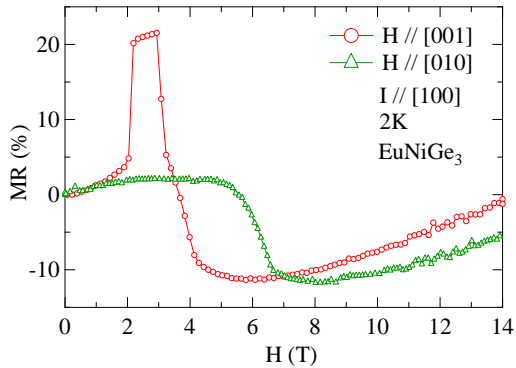}
\caption{\label{fig_MR_comparison_2K}(Color online) Magnetoresistance ratio at 2\,K in EuNiGe$_3$ for $H~\parallel$~[100] (right scale) and $H~\parallel$~[001] (left scale)}
\end{figure}
The vertical fall in MR occurs at the field where the second metamagnetic transition takes place, i.e near 2.8\,T at 1.8\,K. The MR at 2\,K becomes negative close to 4\,T where the magnetization is close to its saturation value and the Eu moments are in the field-induced ferromagnetic state. A negative MR is typically seen in ferromagnets. At 12\,K, the MR is initially positive up to about 1.5\,T and then becomes negative at higher fields. A spin-flop like behavior is seen in the magnetization at 12\,K at around 1.5\,T. At a higher temperature of 20\,K the MR is negative throughout and smaller in absolute magnitude than at 12\,K, signifying the increasing influence of thermally induced scattering on the motion of charge carriers. For  $H\parallel$~[100], the MR field variations up to 14\,T at 2, 5, 8, 12 and 20\,K are shown in Fig.~\ref{fig_MR_100_diff_temperatures}. 
\begin{figure}[h]
\includegraphics[width=0.35\textwidth]{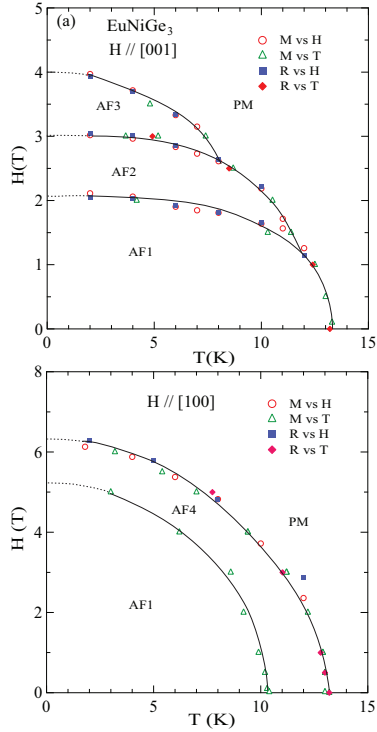}
\caption{\label{fig_Phase_diagram}(Color online) Magnetic phase diagram of EuNiGe$_3$ for (a) $H~\parallel$~[001] and (b) $H~\parallel$~[100]. The label PM is for paramagnetic phase, and the labels AF1, AF2 ... for different AFM phases. Lines are guide to the eyes.}
\end{figure}
At 2\,K, MR is slightly positive up to 5\,T and then begins to decrease and becomes negative close to 6\,T and remaining so up to 14\,T. The sign transition occurs in the field region where the magnetization along [100] shows a minor anomaly followed by the spin-flip transition at 6\,T (see Fig.~\ref{fig_M_vs_H_1p8K_all_directions}). As the temperature increases the field at which the anomaly occurs decreases and the MR also changes sign at lower fields. At 12 and 20\,K, the MR is negative in the entire range of applied fields, as expected in the paramagnetic phase under field. 
Fig.~\ref{fig_MR_comparison_2K} shows a comparison of the MR for $H~\parallel$~[001] and [100] at 2\,K. For fields exceeding 6\,T the MR is comparable along the two directions and hence in conformity with the magnetization data.\\
From the field and temperature dependence of magnetization and electrical resistance, (H,T) phase diagrams have been constructed for the principal crystallographic directions, which are shown in Fig.~\ref{fig_Phase_diagram}. It is observed that the boundaries derived from different  measurements corroborate each other and the dotted lines are possible extrapolations beyond the measurement limits. In the upper panel ($H\parallel$~[001]), three different phases are inferred labeled as AF1, AF2 and AF3 with critical fields of 2, 3 and 4 Tesla, respectively at absolute zero temperature and triple points at (2.6~T, 8~K) and (1.1 T, 12~K). The phase diagram  for $H\parallel$~[100] shown in the bottom panel of the Fig~\ref{fig_Phase_diagram} is different from that of $H\parallel$~[001]. There are no triple points and the critical fields at absolute zero temperature are realatively higher (5.2 and 6.4 Tesla). The phase boundary between AF1 and AF4 possibly separates the commensurate and incommensurate modulated antiferromagnetic structures as indicated by the M\"{o}ssbauer spectra (see section~\ref{mossb}). 
\subsection{M\"{o}ssbauer spectroscopy} \label{mossb}
\begin{figure}[h]
\vspace{0.5cm}
\includegraphics[width=0.45\textwidth]{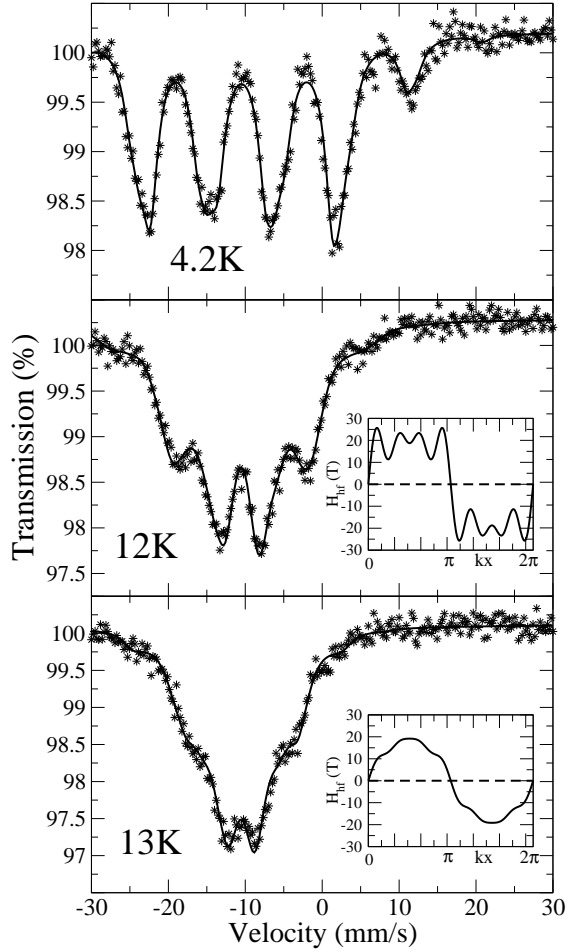}
\caption{\label{fig_Mossbauer_spectra}(Color online) $^{151}$Eu M\"{o}ssbauer absorption spectra at 4.2, 12 and 13\,K in EuNiGe$_3$. For the 4.2\,K spectrum, the line is a fit to a single hyperfine field pattern. For the 12 and 13\,K spectra, the line is a fit to a distribution of hyperfine fields arising from an incommensurate moment modulation. The insets show the corresponding hyperfine field modulations.}
\end{figure}

$^{151}$Eu M\"{o}ssbauer spectra have been recorded at 4.2, 10, 11, 12, 13 and 14\,K in EuNiGe$_3$ (see Fig.~\ref{fig_Mossbauer_spectra}). The spectra below 10\,K show a magnetic hyperfine spectrum due to a single hyperfine field of 30.6(1)\,T at 4.2\,K and 25.7(1)\,T at 10\,K, with an Isomer Shift with respect to Sm$^{*}$F$_3$ of $-$10.6\,mm/s. All these values are characteristic of a divalent Eu ion.
Above 10\,K and up to 13\,K, the spectra change shape and become characteristic of a distribution due to an incommensurate modulation of collinear hyperfine fields.~\cite{eupdsb} At 14\,K, the spectrum is a broad featureless line characteristic of the paramagnetic phase. 

This hyperfine field modulation arises from a modulation of moments since the hyperfine field is proportional to the Eu$^{2+}$ magnetic moment to a good approximation when the former has the standard value of $\simeq$30\,T. The change in spectral shape occurs between 10 and 11\,K, and thus we think that it correlates with the broad anomaly in specific heat and susceptibility observed near T$_{\rm IC}$=10.5\,K. The AFM moment modulation along the direction of the propagation vector {\bf k} was described by the expression:
\begin{equation}
m(kx)\ =\ \sum_{l=0}^4\ m_{2l+1} \sin\ (2l+1)kx
\end{equation}
where the $m_{n}$ are the odd Fourier coefficients of the modulation and are fitted to the shape of the spectrum. The shape of the modulation is represented in the insets of Fig.~\ref{fig_Mossbauer_spectra}. It tends to a sine-wave shape close to T$_{\rm N}$ and squares up when approaching T$_{\rm IC}$. The situation in EuNiGe$_3$ is therefore similar to that in EuPtSi$_3$ \cite{Neeraj EuPtSi3}, where a cascade of transitions is also present, the intermediate phase being an amplitude modulated magnetic phase. One difference is that the specific heat anomaly marking the transition to the commensurate AFM phase is much sharper in EuPtSi$_3$ than in EuNiGe$_3$. One can also notice that the presence of an ICM phase just below T$_{\rm N}$ in the EuMT$_3$ series seems to be correlated with the presence of anisotropy: in EuPtGe$_3$, where the magnetisation is rather isotropic, no ICM phase is present.~\cite{Neeraj EuPtGe3}

\section{Model for computing The single crystal magnetization} \label{model}
In the centered tetragonal lattice of the Eu$^{2+}$ ions (without however inversion symmetry due to different arrangements of the Ni and Ge ions above and below a given plane), one can consider three exchange integrals: nearest neighbour intra-plane integral J$_1$ (ion separation \textit{a}), nearest neighbour interplane integral J$_c$ (ion separation  1/2~$\surd{(c^2+2a^2)}$, both considered in Ref.~\onlinecite{Goetsch} and using the notations therein, to which one can add a next-nearest neighbour integral J$_2$ between ions in two non-adjacent planes (ion separation \textit{c}).~\cite{Herpin} With the lattice parameters in EuNiGe$_3$, these ion separations are respectively about 0.43\,nm, 0.6\,nm and 1\,nm. Thus we restrict ourselves to exchange J$_1$ and J$_c$, and neglect J$_2$ since the corresponding ion separation is much larger. According to the relative signs and magnitudes of J$_1$ and J$_c$ (our convention is a negative integral corresponds to an AFM interaction), the two simplest AFM structures are: \\
(i) ferromagnetic (\textit{ab}) planes with alternating moment directions along c (S1 structure wih \textbf k~=~[001]) \\
(ii) ferromagnetic “stripes” along [110] in the (ab) planes with alternating moment directions in parallel stripes (S2 structure with \textbf k~=~[1/2,1/2,0]).\\
Using the molecular field relations for the N\'eel temperature $T_{\rm N}$ and the paramagnetic Curie temperature $\theta_{\rm p}$ quoted in Ref.~\onlinecite{Goetsch}: 
\begin{eqnarray}
k_B T_{\rm N} & = &\frac{S(S+1)}{3} \sum_{j} J_{ij}\ \cos\phi_{ij} \\
k_B \theta_{\rm p} & = & \frac{S(S+1)}{3} \sum_{j} J_{ij}
\end{eqnarray}
where $\tilde J$ is the exchange tensor, the sum runs over the neighbours of ion i and $\phi_{ij}$ is the angle between moments at sites i and j, one can obtain the exchange integrals for each of the magnetic structures S1 and S2. Using an isotropic mean value $\theta_{\rm p}$~=~4\,K and $T_{\rm N}$~=~13\,K, one obtains \\
\begin{enumerate}
\item {for structure S1: $J_1$~=~0.40\,K and $J_c$~=~-0.11\,K,}\\
\item {for structure S2: $J_1$~=~-0.62\,K and $J_c$~=~0.40\,K.}\\ 
\end{enumerate}
In order to compute the magnetization, and in view of its anisotropic behavior, one must add two interactions: \\
(i) the single ion crystalline anisotropy E$_{CF}$, which is small for Eu$^{2+}$ but plays a role in defining the easy plane/axis. Since the local symmetry at the Eu site is tetragonal (\textit{4mm}), we consider only the leading second order term E$_{CF}$~=~DS$_z^2$, where D is a parameter and \textit{Oz} the local fourfold axis, i.e. the \textit c crystal axis. Usually, $\vert$D$\vert$ amounts to a few 0.05\,K. For negative D, the easy axis is the fourfold \textit{c} axis, while for positive D the \textit{ab}-plane is the easy plane.\\
(ii) the standard dipole-dipole interaction $\cal H_{\rm dip}$ which is taken to be of infinite range (see Appendix \ref{dip} for details about the summation over lattice sites.~\cite{Ewald, Wang}) This interaction is also weak, but it is of the same order as the crystalline anisotropy and must be taken into account. The importance of the dipole-dipole interaction in determining magnetic structures for Gd compounds has been examined, using a different approach, in Ref.~\onlinecite{Rotter}. An example of magnetization single crystal curves computed with and without the dipolar interaction is provided in Appendix \ref{dip}. The total hamiltonian acting on an Eu ion is thus:
\begin{equation}
{\cal H} = DS_z^2 + g\mu_{\rm B}\  {\bf H}.{\bf S} -{\bf S}.\sum_j J_{ij} {\bf S}_j + g\mu_{\rm B}\  {\bf S}. {\bf H}_{dip},
\end{equation}
where the different terms are, from left to right, the crystal field, the Zeeman interaction ($g$=2 for Eu$^{2+}$), the exchange coupling restricted to intra-plane and interplane nearest neighbours, and the infinite range dipolar interaction with the field {\bf H}$_{dip}$ acting on the given ion. 

\begin{figure}[h]
\includegraphics[width=0.45\textwidth]{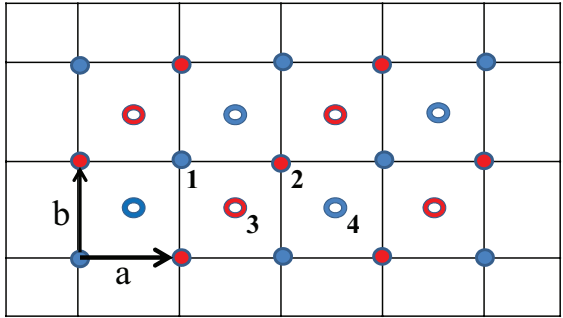}
\caption{\label{fig_Decomposition_into_sublattices}(Color online) Decomposition of the EuNiGe$_3$ tetragonal lattice into 4 sublattices, in projection onto the (ab) plane. The full circles represent Eu atoms in the z~=~0 plane, the open circles those in the z~=~c/2 plane.}
\end{figure}

\begin{figure}[h]
\includegraphics[width=0.45\textwidth]{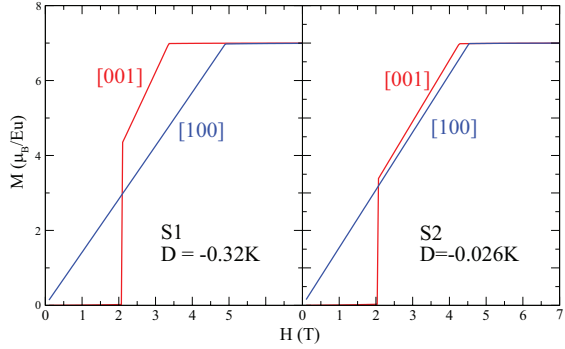}
\caption{\label{fig_Calculated_magnetization_of_S1_and_S2_structures}(Color online) Calculated magnetization curves at 1.8\,K for the S1 (planes) and S2 (stripes) structures along [001] and [100] with isotropic exchange and dipolar interaction. The $\vert$D$\vert$ value is more than 10 times larger for S1 than for S2 in order to obtain a spin-flop field of 2\,T.}
\end{figure}

The sublattice decomposition of the zero-field magnetic structure is essential for the calculation of the single crystal magnetization. In the absence of knowledge of this magnetic structure, and in order to be able to examine the two simple types of aforementioned structures S1 and S2 with the same formalism, we have used a decomposition into 4 sublattices, as sketched in Fig.~\ref{fig_Decomposition_into_sublattices}. When describing an S1 type structure, sublattices 1 and 2 on the one hand, 3 and 4 on the other hand, are identical. As to the S2 structure, sublattices 1 and 4 on the one hand, 2 and 3 on the other hand, are identical. The computation of the magnetization is performed treating the exchange and dipole-dipole interactions in mean field involving the 4 sublattices, in a self-consistent way.
We have first simulated the magnetization curves at 1.8\,K for the S1 and S2 structures relevant to EuNiGe$_3$ using the aforementioned exchange integrals. The value of the D parameter was taken negative and adjusted so that the spin-flop field value, for $H~\parallel$~[001], be 2\,T, as experimentally observed. For the S1 structure, $\vert$D$\vert$ has to be taken rather large lest the structure lies in the (\textit{ab}) plane, which is an effect of the dipolar field. The calculated magnetization curves along [100] and [001] are represented in Fig.~\ref{fig_Calculated_magnetization_of_S1_and_S2_structures}.
%
The obtained spin-flip fields along [001] and [100], 3.3 and 4.8\,T, respectively, for S1, and close to 4.5\,T for S2, do not precisely match the experimental values 4.1 and 6.2\,T, respectively (see Fig.~\ref{fig_M_vs_H_1p8K_all_directions}), and neither the initial non-zero slope of M(H) nor its staircase-like behavior when $H~\parallel$~[001] are  reproduced. For the S1 case, the large difference in spin-flip fields is due to the large value of $\vert$D$\vert$.

\begin{figure}[h]
\includegraphics[width=0.45\textwidth]{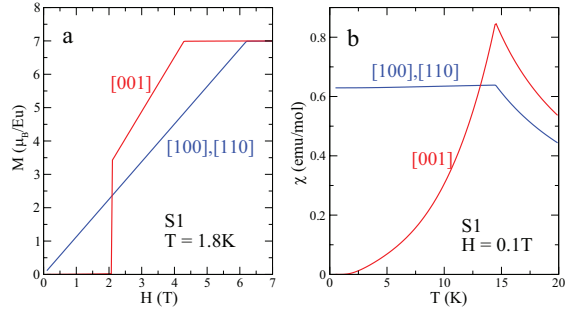}
\caption{\label{fig_Calculated_M_and_Chi_for_S1_structure}(Color online) Calculated magnetization curves at T~=~1.8\,K (a) and susceptibility curves at H~=~0.1\,T (b) for the S1 structure in the presence of exchange anisotropy and dipolar interaction.}
\end{figure}

Another type of possible anisotropy is that of the exchange integrals (see Appendix \ref{exanis} for details). In a simple model, we assume that, for each Eu-Eu bond, the symmetric part of the exchange tensor is diagonal and has axial symmetry in a frame linked with the Eu-Eu bond, i.e. it possesses two components $J^\parallel$ and $J^\perp$. Since we consider the nearest neighbor intraplane and interplane integrals, we introduce the four parameters, $J_1^\parallel$, $J_1^\perp$, $J_c^\parallel$ and $J_c^\perp$. For each type of AFM structure (S1 and S2), it is then possible to find a set of exchange integrals and D value which yields spin-flip field values closer to experiment, and susceptibility curves in reasonable qualitative agreement with the data. These parameter sets are:
\begin{enumerate}
\item {for the S1 structure: $J_1^\parallel$~=~0.23\,K, $J_1^\perp$~=~0.485\,K, $J_c^\parallel$~=~$-$0.125\,K, $J_c^\perp$~=~$-$0.145\,K, and D~=~$-$0.06\,K,}
\item {for the S2 structure: $J_1^\parallel$~=~$-$0.70\,K, $J_1^\perp$~=~$-$0.64\,K, $J_c^\parallel$~=~0.45\,K, $J_c^\perp$~=~0.36\,K, and D~=~$-$0.10\,K.}
\end{enumerate}
The M(H) and $\chi$(T) curves are the same for both sets; the AFM transition temperature is T$_N$~=~14\,K and the paramagnetic Curie temperatures are $\theta_{\rm p}$[100] = $\theta_{\rm p}$[110] = 2.2\,K and $\theta_p$[001]=5.1\,K. The curves corresponding to S1 are shown in Fig.\ref{fig_Calculated_M_and_Chi_for_S1_structure} and should be compared with Figs.~\ref{Chi_vs_T_low_temp_all_directions} and \ref{fig_M_vs_H_1p8K_all_directions}. For the S1 structure, with the introduction of exchange anisotropy, one recovers a smaller, hence more physical, $\vert D\vert$ value. The calculated values for $T_{\rm N}$ and for the $\theta_{\rm p}$'s are rather close to the experimental values, particularly for $H~\parallel$~[001]. The larger value of $\theta_{\rm p}$[001] is an effect of the anisotropy and it is reflected in the larger value of $\chi$(T$_{\rm N}$) for this field direction.  
%
\vspace{0.5cm}
However, it is still not possible to obtain a non-zero initial slope, nor a staircase like behavior, for \textit M(H) along [001] since the easy axis in both S1 and S2 structures is the c axis. This suggests that the zero field structure is a ``canted'' structure, i.e. it possesses a spin component in the (ab) plane, and one must find mechanisms able to induce such a deviation from the easy axis. This is the subject of discussion in the next section.
\section{Discussion}
One can readily think of two mechanisms leading to a magnetic structure canted off the c axis. The first one is a crystallographic distortion that would occur in the AFM phase, such that the Eu site no longer has fourfold point symmetry. This would imply a loss of the isotropy of the magnetic properties in the (ab) plane, which contradicts observations since the magnetization and susceptibilty along [100] and [110] are identical. So a low temperature symmetry breaking distortion can be discarded.

The second mechanism is the presence of an asymmetric term of the exchange tensor, namely the Dzyaloshinski-Moriya interaction. It can be seen to be non-zero for the two exchange paths considered here since none of the mid points of the intraplane and interplane exchange bonds is an inversion center for the crystal structure of EuNiGe$_3$. Considering only nearest neighbour exchange, i.e. intraplane exchange, application of the Moriya rules~\cite{Moriya} shows that the associated Dzyaloshinski-Moriya vector D$_{DM}$ lies in the (ab) plane and is perpendicular to the Eu-Eu bond. This can lead to a canted zero field magnetic structure. However, in order to test this hypothesis, a more sophisticated sublattice decomposition is needed, because convergence cannot be reached in our present in-field calculation in the presence of asymmetric exchange. This suggests that the 4 sublattices used here are not adequate to reproduce the actual magnetic structure and its evolution with the field, and one must await the determination of the magnetic structure of EuNiGe$_3$ to check for the presence and magnitude of the Dzyaloshinski-Moriya exchange. 

One can also note that the absence of mirror symmetry with respect to the (ab) plane implies that there should be two different interplane nearest neighbour exchange integrals $J^u_c$ and $J^l_c$, according to the upper or lower position of the Eu-Eu bond with respect to the (ab) plane. This asymmetry is a priori not expected to lead to a transverse spin component since it does not break the fourfold symmetry around the c axis. With our present definition of the 4 sublattices, which are invariant by a translation of vector c, introduction of these different integrals amounts to replacing $J_c$ by ½($J^u_c$+ $J^l_c$) and therefore this kind of asymmetry is not relevant within our model.

Finally, an anisotropic magnetic behavior has also been observed in EuPtSi$_3$~\cite{Neeraj EuPtSi3} and, for the case of the other S-state ion Gd$^{3+}$, in several intermetallics: GdRu$_2$T$_2$ (T~=~Ge or Si),~\cite{Garnier} Gd$_2$PdSi$_3$~\cite{Saha} and Gd$_5$Ge$_4$,~\cite{Ouyang} where a metamagnetic behavior of the magnetization is reported along some particular direction. This behavior was attributed in these works to exchange anisotropy. 
\section{Conclusion}
As a conclusion, single crystals of the intermetallic material EuNiGe$_3$ have been synthesized and studied with a variety of techniques: magnetization, specific heat, transport and M\"{o}ssbauer spectroscopy. EuNiGe$_3$ contains divalent Eu, its N\'eel temperature is found to be T$_N$~=~13.2\,K, in agreement with Ref.~\onlinecite{Goetsch}. The M\"ossbauer spectra show that there occurs a cascade of transitions, as often observed in intermetallic Eu materials, from paramagnetic to incommensurate AFM to commensurate AFM as temperature decreases. The magnetization and magnetoresistance in the AFM phase are found to be strongly anisotropic, which is a priori surprising for a spin only ion like Eu$^{2+}$. In particular, the magnetization along the c axis at 2\,K shows an unusual staircase-like behavior. With the aim of obtaining the exchange integrals in EuNiGe$_3$, we applied a model with anisotropic exchange and dipole-dipole interactions which reproduces the main features of the magnetization and susceptibility curves. However, we could not reach a fully satisfying reproduction of the magnetization curves along c, which suggests that the magnetic structure is more complex than the AFM collinear arrangement assumed in our model. To go further, one must await the determination of the magnetic structure by neutron diffraction. We emphasize that the dipole-dipole interaction must be taken into account for a full understanding of the magnetic structure and its evolution with the applied field in Eu intermetallics.

\appendix 

\section{The dipole-dipole interaction} \label{dip}

We show here that it is important to take into account the dipole-dipole interaction in computing the single crystal magnetization curves. Indeed, Eu$^{2+}$ has a rather large saturated moment (7\,$\mu_B$), and the dipolar field is of the order of 0.05\,T. This value is small with respect to the characteristic exchange fields in EuNiGe$_3$ (a few T), but it is of the same magnitude as the anisotropy field in Eu$^{2+}$ compounds. Therefore, it must be taken into account for determining both the zero field magnetic structure, since it can compete with the crystalline anisotropy in defining the easy magnetic axis, and the low field magnetization since it plays a role in defining the spin-flop field.

In order to obtain the dipolar field on a given Eu$^{2+}$ ion on a given sublattice, we use the Ewald summation method to compute the infinite lattice sums \cite{Ewald,Wang}, and we write the dipolar field acting on an ion in the $l$ sublattice as:
\begin{equation}
{\bf H}_{dip}^l=\displaystyle\sum_k Q_k^l\  {\bf m}_k
\end{equation}
where the $Q_k^l$ matrices contain lattice sums. For instance, the self-sublattice matrix $Q_l^l$ is diagonal and writes, in units of the Lorentz field parameter H$_L$~=~$\frac{4\pi}{3V} \mu_B$, where $V$~=~2\textit{a}$^2$c is the volume of the primitive cell, and \textit{a}~=~0.4388\,nm and \textit{c}~=~0.9895\,nm:
\begin{eqnarray}
Q_l^l = \left( \begin{array}{ccc}
1.7345 & 0 & 0 \\
0 & 1.7345 & 0 \\
0 & 0 & -0.4690 \end{array} \right) 
\end{eqnarray}
and the matrix $Q_{3}^{1}$ writes, in the same units:
\begin{eqnarray}
Q_{3}^{1} = \left( \begin{array}{ccc}
0.0620 & -0.3822 & 0 \\
-0.3822 & 0.0620 & 0 \\
0 & 0 & 2.8760 \end{array} \right). 
\end{eqnarray}
\begin{figure}[h]
\includegraphics[width=0.45\textwidth]{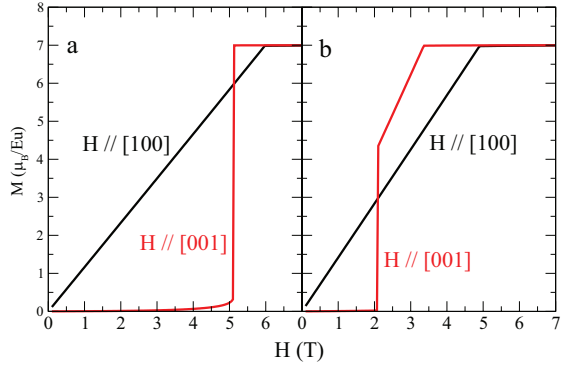}
\caption{\label{fig_M_in_presence_and_absence_of_dipolar_field}(Color online) Self-consistent computation of the magnetization curves at 1.8\,K for $J_1$~= 0.40\,K, $J_c$~=~ $-$0.11\,K and D~=~$-$0.32\,K, for two directions of the applied magnetic field, in the absence (a) and in the presence (b) of the dipolar field.}
\end{figure}
To illustrate the importance of taking into account the dipolar interaction, we consider the S1 magnetic structure (ferromagnetic planes), with J$_1$~=~0.40\,K and J$_c$~=~$-$0.11\,K. For D~$<$~0, in the absence of dipolar field, the Eu moments lie along the \textit c axis. However, when taking into account the dipolar field, the Eu moments lie in the \textit{ab}-plane until D reaches the rather large value $-$0.23\,K, and there is no spin-flop for $H~\parallel$~[001]. For D~$<$~$-$0.23\,K, the easy c axis is recovered and a spin-flop occurs for $H~\parallel$~[001], and as $\vert D\vert$ increases the spin-flop field increases. As an illustration of the importance of the dipole-dipole interaction for the calculation of the magnetization curves, we have represented in Fig.~\ref{fig_M_in_presence_and_absence_of_dipolar_field} the M(H) curves for the S1 magnetic structure without and with the dipolar field with the same (probably unphysically large) value of the D parameter~:~D~=~$-$0.32\,K. In this case, introduction of the dipolar field has the effect of decreasing the spin-flop field, which in turns affects the value of the spin-flip field for $H~\parallel$~[001]. With this example, we want to show that the dipolar field should be taken into account for any realistic calculation of the single crystal magnetization in Eu compounds, especially when the ordering temperature is not too high, i.e. the exchange fields are not too large with respect to the dipolar field ($\approx$~0.05\,T).

\section{The anisotropic exchange tensor} \label{exanis}

We consider a simple implementation of anisotropy of the exchange interaction, by considering that the exchange tensor J$_{ij}$ has axial symmetry in a frame where the Eu-Eu bond is taken as the \textit{z}-axis, i.e. it has a longitudinal component J$^{\parallel}$ and two identical transverse components J$^\perp$. Then, for each considered exchange bond $\alpha$ (intraplane or interplane), the exchange Hamiltonian writes, in a frame linked with the Eu-Eu bond:
\begin{equation}
{\cal H}_{\rm ex} = -J_\alpha^{\parallel}S_i^z S_j^z - J_\alpha^\perp(S_i^xS_j^x+S_i^yS_j^y).
\end{equation}
For instance, considering the intraplane exchange bond between sublattice 1 and 2 along a (see Fig.~\ref{fig_Decomposition_into_sublattices}), the frame Oxyz linked to this bond is taken as: z~$\parallel$~a, x~$\parallel$~c and y~$\parallel$~-b. The diagonal axial exchange tensor with components $J_1^{\perp}$ and $J_1^{\parallel}$ is then transformed into the (abc) frame and, taking into account 4 such intraplane bonds, one obtains the total intraplane exchange matrix between sublattices 1 and 2, or 3 and 4, to which one can add an antisymmetric Dzyaloshinski-Moriya term with vector \textbf{D}$_{DM}$ which lies in the \textit{ab}-plane and is perpendicular to the bond: 
\begin{eqnarray}
 J_1^t = 2\left( \begin{array}{ccc}
J_1^\perp+J_1^\parallel & 0 & -D_{DM} \\
0 & J_1^\perp+J_1^\parallel & -D_{DM} \\
D_{DM} & D_{DM} & 2J_1^\perp \end{array} \right).
\end{eqnarray} 
Repeating this procedure for the interplane exchange (without Dzyaloshinski-Moriya term), one finds that the total interplane exchange matrix between sublattices 1 and 3 or 1 and 4, 2 and 3 or 2 and 4, reads:
\begin{eqnarray}
J_c^t = 2\left( \begin{array}{ccc}
J_c^\perp+\delta & \delta & 0 \\
\delta & J_c^\perp+\delta & 0 \\
0 & 0 & J_c^\parallel-2\delta \end{array} \right),
\end{eqnarray}
where $\delta = \frac{J_c^\parallel-J_c^\perp}{2+y^2}$ and y~=~c/a.
\end{document}